\documentclass[12pt,aps]{revtex4}
\usepackage{times,mathptm}
\usepackage{amsmath,amsfonts,bm}
\usepackage{graphicx}

\begin{document}

\title{The error-floor of LDPC codes in the Laplacian channel
%\footnote{This manuscript is submitted for the Forty-Third Annual
%Allerton Conference on Communication, Control, and Computing to be
%held in Allerton, IL September 28-30, 2005.}
}

\author{M.~G.~Stepanov ${}^{1,2}$ and M.~Chertkov ${}^1$}

\affiliation{${}^1$ Complex Systems Group and Center for Nonlinear
Studies, Theoretical Division,  Los Alamos National Laboratory, Los
Alamos,
 NM 87545, USA\\
 ${}^2$ Institute of Automation and Electrometry,
Novosibirsk 630090, Russia \\ E-mails: stepanov@cnls.lanl.gov,
chertkov@lanl.gov}

%\date{\today}

\begin{abstract}

We analyze the performance of Low-Density-Parity-Check codes in the
error-floor domain where the Signal-to-Noise-Ratio, $s$, is large,
$s\gg 1$. We describe how the instanton method of theoretical
physics, recently adapted to coding theory, solves the problem of
characterizing the error-floor domain in the Laplacian channel. An
example of the $(155,64,20)$ LDPC code with four iterations (each
iteration consisting of two semi-steps: from bits-to-checks and from
checks-to-bits) of the min-sum decoding is discussed. A generalized
computational tree analysis is devised to explain the rational
structure of the leading instantons. The asymptotic for the symbol
Bit-Error-Rate in the error-floor domain is comprised of individual
instanton contributions, each estimated as $\sim\exp(-l_{\rm inst;L}
\cdot s)$, where the effective distances, $l_{\rm inst;L}$, of the
the leading instantons are $7.6, 8.0$ and $8.0$ respectively. (The
Hamming distance of the code is $20$.) The analysis shows that the
instantons are distinctly different from the ones found for the same
coding/decoding scheme performing over the Gaussian channel. We
validate instanton results against direct simulations and offer an
explanation for remarkable performance of the instanton
approximation not only in the extremal, $s\to \infty$, limit but
also at the moderate $s$ values of practical interest.

\end{abstract}

\maketitle

A novel exciting era has begun in coding theory with the discovery of
Low-Density-Parity-Check (LDPC) \cite{63Gal,99Mac} and turbo
\cite{93BGT} codes. These codes are special, not only because they can
approach the virtually error-free transmission limit, but mainly because
a computationally efficient iterative decoding scheme is readily
available. When operating at moderate noise values these approximate
decoding algorithms show an unprecedented ability to correct errors, a
remarkable feature that has attracted a lot of attention
\cite{96Wib,99FKKR,04Ric}. However, it was also shown that the
approximate algorithms are incapable of matching the performance of
Maximum-Likelihood (ML) decoding beyond the so-called error-floor
threshold found at higher Signal-to-Noise-Ratios (SNR). The importance
of error-floor, i.e. highest SNR, analysis was recognized in the early
stages of the turbo code revolution \cite{96BM}, and it soon became
apparent that LDPC codes are also not immune to the error-floor
deficiency \cite{01MB,04Ric}. To estimate the error-floor asymptotics in
modern high-quality systems is a notoriously difficult task because
direct numerical methods, e.g.\ Monte Carlo, cannot be used to determine
Bit-Error-Rate (BER) below $10^{-9}$. The main approaches to the
error-floor analysis problem proposed to date include: (i) a heuristic
approach of the importance sampling type \cite{04Ric}, utilizing
theoretical considerations developed for a typical randomly constructed
LDPC code performing over the binary-erasure channel \cite{02DPTRU}, and
(ii) deriving lower bounds for BER \cite{04VK}.

Recently, we (in collaboration with V.~Chernyak and B.~Vasic) have
also proposed a physics inspired approach that is capable of a
computationally tractable analysis of the error floor phenomenon
\cite{05SCCV}. An efficient numerical scheme was proposed, which was
ab-initio by construction, i.e.\ the optimization scheme required no
additional assumptions (e.g.\ no sampling). This numerical
optimization scheme, called the instanton-amoeba scheme, which plays
a central role in our analysis, was shown to be accurate at
producing configurations whose validity, for actual optimal noise
configurations, can be verified theoretically. Finally, the
instanton-amoeba scheme, introduced in \cite{04CCSV}, is also
generic, in that there are no restrictions related to the type of
decoding or the channel.

To illustrate the last,  most important point,  we complement the
analysis of \cite{05SCCV} that focused on the white Gaussian
symmetric channel, by using the new results for the white Laplacian
channel explained in this manuscript. Our choice of the Laplacian
channel is arbitrary,  e.g. influenced by its relevance to the
description of fiber-optics communication channels. (Amplifier
noise, known to have Gaussian statistics, contributes additively to
the electric signal carrying the information. However, the
intensity, the electric field squared, is detected on the receiver
end. Therefore, the resulting transition probability, characterizing
the noise in the fiber optics channel as the whole, shows
exponential tails. See \cite{97Agr} for a discussion of these and
other statistical errors in fiber optics channels.)

Our goal is to demonstrate using the example of the Laplacian
channel that
\begin{itemize}
\item{The numerical optimization approach to finding the most damaging
configuration of the noise, instanton-amoeba, is computationally
efficient.}
\item{Subsequent theoretical analysis of the iterative decoding instanton
based on the notion of the computational tree is channel specific,
but it is also generalizable, i.e. the theoretical scheme can be
modified to explain the rational structure of instantons for any, in
particular Laplacian, channel.}
\item{Instanton configurations for different channels are different.
Thus, information on the error-floor analysis available for a
channel (say, for the Gaussian channel) does not allow quantitative
description of the error-floor in another channel (say, in the
Laplacian channel).}
\end{itemize}

This manuscript is organized as follows. We describe the problem of
error-floor analysis, introduce the Laplacian channel and briefly review
the results of \cite{05SCCV} in Section~\ref{sec:intro}. Our theoretical
analysis, generalizing the computational tree approach of \cite{96Wib},
is detailed in Section~\ref{sec:theo}. Section~\ref{sec:amoeba} explains
results of numerical instanton-amoeba evaluation.  Here we present the
three most important (for $s\gg 1$) instantons, for example of the
min-sum decoding with four iterations performing over the $(155,64,20)$
LDPC code described in \cite{01TSF}. We also discuss here how the
numerically found instanton is rationalized/explained theoretically. The
comparison of Monte Carlo simulations and instanton prediction is
discussed in Section~\ref{sec:moderate}. Some final remarks and comments
are presented in Section~\ref{sec:concl} concluding the manuscript.

\section{Setting the problem}
\label{sec:intro}

Let us begin by introducing notation. A message word consisting of
$K$ bits is encoded in an $N$-bit long codeword, $N>K$. In the case
of binary, linear coding a convenient representation of the code is
given by $M\geq N-K$ constraints, often called parity checks or
simply checks. Formally, ${\bm\sigma} = (\sigma_{1}, \dots,
\sigma_{N})$ with $\sigma_{i} = \pm 1$, is one of the $2^{K}$
codewords if and only if $\prod_{i\in \alpha } \sigma_{i} = 1$ for
all checks $\alpha = 1, \dots, M$, where $i\in\alpha$ if the bit $i$
contributes the check $\alpha$. The relation between bits and checks
(we use $i\in\alpha$ and $\alpha\ni i$ interchangeably) is often
described in terms of the $M\times N$ parity-check matrix $\hat{ H}$
consisting of ones and zeros: $H_{\alpha i}=1$ if $i\in\alpha$ and
$H_{\alpha i}=0$ otherwise. A bipartite graph representation of
$\hat{H}$, with bits marked as circles, checks marked as squares,
and edges corresponding to the respective nonzero elements of
$\hat{H}$, is called the Tanner graph of the code. For an LDPC code
$\hat{H}$ is sparse, i.e.\ most of the entries are zeros.
Transmitted through a noisy channel, a codeword gets corrupted due
to the channel noise, so that the channel output is ${\bm
x}\neq{\bm\sigma}$. Even though information about the original
codeword is lost at the receiver, one still possesses the full
probabilistic information about the channel, i.e.\ the conditional
probability, $P({\bm x}|{\bm \sigma}')$, for a codeword ${\bm
\sigma}'$ to be a pre-image for the output word ${\bm x}$, is known.
In the case of independent noise samples, the full conditional
probability can be decomposed into the product, $P({\bm x}|{\bm
\sigma}')=\prod_i p(x_i|\sigma_i')$. A convenient characteristic of
the channel output at a bit is the so-called log-likelihood,
$h_i=\log[p(x_i|+1)/p(x_i|-1)]/2$. The decoding goal is to infer the
original message from the received output, ${\bm x}$. ML decoding
(which generally requires an exponentially large number, $2^K$, of
steps) corresponds to finding the most probable transmitted codeword
given ${\bm x}$. Belief Propagation (BP) decoding \cite{63Gal,99Mac}
constitutes a fast (linear in $K, N$) yet generally approximate
alternative to ML. As shown in \cite{63Gal} the set of equations
describing BP becomes exactly equivalent to the so-called symbol
Maximum-A-Posteriori (MAP) decoding in the loop-free approximation
(a similar construction in physics is known as the Bethe-tree
approximation \cite{35Bet}), while in the low-noise limit, i.e. in
the limit of very large SNR, $s\to\infty$, ML and MAP become
indistinguishable and the BP algorithm reduces to the min-sum
algorithm:
\begin{equation}
  \eta^{(n+1)}_{i\alpha} = h_i + \sum_{\beta\neq\alpha}^{\beta\ni i}
  \prod_{j\neq i}^{j\in\beta} \mbox{sign}
    \big[ \eta^{(n)}_{j\beta} \big] \min_{j\neq i}^{j\in\beta} \big|
    \eta^{(n)}_{j\beta} \big|,
  \label{min_sum}
\end{equation}
where the message field $\eta^{(n)}_{i\alpha}$ is defined on the
edge that connects bit $i$ and check $\alpha$ at the $n$-th step of
the iterative procedure and $\eta^{(0)}_{i\alpha} \equiv 0$. The
result of decoding is determined by a-posteriori log-likelihood,
$m^{(n)}_i$, defined by the right-hand-side of Eq.~(\ref{min_sum})
with the restriction $\beta\neq\alpha$ dropped.  The BER measuring
probability of errors at a given bit $i$ becomes
\begin{equation}
  B_{i}=\int d{\bm x}\ \theta\!\left(-m_{i}\{\bm x\} \right)
    P({\bm x}|{\bm 1}),
  \label{BER}
\end{equation}
where $\theta(z)=1$ if $z>0$ and $\theta(z)=0$ otherwise;
${\bm\sigma} = {\bm 1}$ is assumed for the input (since in a
symmetric channel the BER is invariant with respect to the choice of
the input codeword). When the BER is small (SNR is large) the
integral over output configurations ${\bm x}$ in Eq.~(\ref{BER}) is
approximated by:
\begin{eqnarray}
B_i\sim \sum_{\rm inst} V_{\rm inst} \times P({\bm x}_{\rm inst}|{\bm 1}),
\label{Bi}
\end{eqnarray}
where ${\bm x}_{\rm inst}$ are the special instanton configurations
of the output maximizing $P({\bm x}|{\bm 1})$ under the
error-surface condition, $m_i\{{\bm x}\}=0$; $V_{\rm inst}$ combines
combinatorial and phase-volume factors (the latter one accounts for
what physicists call fluctuations around the respective instanton).
Individual instanton contributions into the rhs of Eq.~(\ref{Bi})
decrease significantly with increasing SNR. Thus at large SNR only
instanton(s) with the highest $P({\bm x}_{\rm inst}|{\bm 1})$ is
(are) relevant.

For the common model of the Additive White Gaussian Noise (AWGN)
channel,
\begin{eqnarray}
p_{\rm G}(x|\sigma) = \exp(-s^2 (x - \sigma)^2/2)/\sqrt{2\pi/s^2},
\label{gauss}
\end{eqnarray}
finding the instanton, ${\bm \xi}_{\rm inst} = {\bm 1} - {\bm
x}_{\rm inst} \equiv l({\bm u}){\bm u}$, turns into minimizing the
length $l({\bm u})$ with respect to the unit vector in the noise
space ${\bm u}$, where $l({\bm u})$ measures the distance from the
zero-noise point to the point on the error surface corresponding to
${\bm u}$. This task of the instanton analysis for the AWGN channel
was discussed in detail in \cite{05SCCV}. In \cite{05SCCV} we
developed a numerical scheme where the value of the length $l({\bm
u})$ for any given unit vector ${\bm u}$ was found by the bisection
method. The minimum of $l({\bm u})$ was found by a downhill simplex
method also called ``amoeba'' \cite{88Pre}, with accurately tailored
(for better convergence) annealing. (Note, that even previously, the
numerical instanton method was successfully verified in
\cite{04CCSV} against analytical results in the loop-free case.) To
demonstrate the utility of this method we chose in \cite{05SCCV} the
example of the $(155,64,20)$ LDPC code of \cite{01TSF}. (The parity
check matrix of the code is shown in Fig.~S1 of \cite{05SCCV}.) The
code includes $155$ bits and $93$ checks. Each bit is connected to
three checks while any check is connected to five bits. The minimal
Hamming distance of the code is $l_{\rm ML;G}^2 = 20$, i.e.\ for
$s\gg 1$, and if the decoding is ML, BER becomes $\sim \exp(-20 \cdot
s^2/2)$ in the Gaussian channel. Iterative decoding is suboptimal,
thus respective error-floor asymptotics become $P({\bm x}_{\rm
inst}|{\bm 1})\sim \exp(-l_{\rm inst;G}^2 \cdot s^2/2)$. The numerical,
and subsequent theoretical, analyses of \cite{05SCCV} suggest that
the instantons, as well as the respective effective distance,
$l_{\rm inst;G}$, do depend on the number of iterations. Focusing
primarily on the already nontrivial case of four iterations, we
showed in \cite{05SCCV} that the three minimal weight (largest
probability) instantons have effective lengths, $l_{\rm a;G}^2 =
46^2/210 \approx 10.076$, $l_{\rm b;G}^2 = 806/79 \approx 10.203$
and $l_{\rm c;G}^2 = 44^2/188 \approx 10.298$ respectively. These
instantons were found as the result of multiple attempts at
instanton-amoeba minimization. The remarkable integer/rational
structure of the instantons found numerically by instanton-amoeba
for the AWGN channel admits a clear theoretical explanation
discussed in details in \cite{05SCCV}. We have already suggested in
\cite{05SCCV} that the instanton analysis (in both numerical and
theoretical parts) is actually generic, and is thus applicable to
wide range of different codes and channels.

To illustrate this last point we focus in this manuscript on
analysis of the generalized Additive White Laplacian Noise (AWLN)
channel:
\begin{eqnarray}
  p_{\rm GLap}(x|\sigma) \propto \exp
    \left(-s\sqrt{(x-\sigma)^2+\alpha^2} \right), \label{Lap1}
\end{eqnarray}
where $s$ is the signal-to-noise-ratio (SNR) and $\alpha$ is the
regularization parameter. (We are mainly interested in the
$\alpha\to 0$,  thus $\alpha$ is introduced primarily for the
purpose of accurately regularizing/resolving the singularity at
$\xi=x-\sigma=0$.) If the detected signal at a bit is $1-\xi$, the
respective log-likelihood at the bit is defined as
\begin{equation}
 h=\frac{1}{2s}\ln\left[\frac{p(\xi)}{p(2-\xi)}\right]=
 \Biggl.\frac{\sqrt{(2-\xi)^2+\alpha^2}-\sqrt{\xi^2+\alpha^2}}{2}
 \Biggr|_{\alpha\to 0}\to
 \Biggl\{ \begin{array}{cc} -1, & \xi\geq 2;
 \\ 1-\xi, & 2\geq \xi\geq 0;\\ +1, & 0\geq \xi.\end{array}\Biggr. ,
 \label{h1}
\end{equation}
where one chooses to measure log-likelihoods in the AWLN channel in
units of SNR,  $s$ (and not in units of the SNR squared that was
natural choice in the AWGN channel).

In the Laplacian channel ($\alpha = 0$), finding the instanton means to
minimize $l({\bm u})\sum_i |u_i|$ with respect to the unit vector ${\bm
u}$ in the $N$-dimensional space, rather then minimizing $l({\bm u})$
that was the case for the Gaussian channel. Notice, that just from this
fact one finds absolutely no reason to expect that instantons for
Gaussian and Laplacian channels are in any way related to each other.
The contribution of an instanton to the BER for $s\gg 1$, characterized
by the effective distance $l_{\rm inst;L}$, is estimated as $\sim
\exp(-l_{\rm inst;L} \cdot s)$ in the case of the Laplacian channel.

\section{Generalized computational tree analysis in the Laplacian channel}
\label{sec:theo}

In this Section we describe a theoretical approach to the instanton
analysis. This analysis, based on the computational tree
construction of Wiberg \cite{96Wib}, was discussed in \cite{05SCCV}
for the Gaussian channel. Here we explain how the analysis can be
modified to describe instantons in the case of the AWLN channel.

We start with the universal,  i.e. channel insensitive, part of the
construction. The computational tree is built by unwrapping the
Tanner graph of a given code into a tree from a bit for which we
would like to determine the probability of error. (This bit will be
called erroneous bit.) The number of generations in the tree is
equal to the number of min-sum iterations. As observed in
\cite{96Wib}, the result of decoding at the erroneous bit of the
original code is exactly equal to the decoding result in the tree
center. It should be noted that once log-likelihoods representing an
instanton are distributed on the tree, one can verify directly (by
propagating messages from the leaves to the tree center) that the
algorithm produces zero a-posteriori log-likelihood at the tree
center. Any check node processes messages coming from the tree
periphery in the following way: (i) the message with the smallest
absolute value (one assumes no degeneracy) is passed, (ii) the
source bit of the smallest message is ``colored'', and (iii) the
sign of the product of inputs is assigned to the outcome. At any bit
that lies on the colored leaves-to-center path the incoming messages
are summed up. The initial messages at any bit of the tree are
log-likelihoods and, therefore, the result obtained in the tree
center is a linear combination of the log-likelihoods with integer
coefficients. The integer $n_i$ corresponding to bit $i$ of the
original graph is the sum of the signatures over all colored
replicas of $i$ on the computational tree. Therefore the condition
at the tree center becomes $\sum_i n_i h_i = 0$.

So far the discussion has been generic. Let us now adapt this
generic construction to the case of the generalized AWLN channel
described by Eq.~(\ref{Lap1}). Returning from the computational tree
to the original graph and maximizing the integrand of
Eq.~(\ref{BER}) with the condition, $\sum_i n_i h_i = 0$, enforced
we arrive at the following expressions for the effective length:
\begin{eqnarray}
  l_{\rm inst;L}=\sum_i \sqrt{\xi_i^2+\alpha^2} + \lambda \sum_i n_i
    h_i, \label{S1}
\end{eqnarray}
where $\lambda$ is the Lagrange multiplier enforcing the zero
a-posteriori log-likelihood condition at the tree center. Minimizing
Eq.~(\ref{S1}) with respect to $\xi_i$ one derives
\begin{eqnarray}
  \frac{\xi_i}{2-\xi_i} \sqrt{\frac{(2-\xi_i)^2+\alpha^2}
    {\xi^2_i+\alpha^2}} = \frac{\lambda n_i}{2-\lambda n_i}.
  \label{var1}
\end{eqnarray}
Expressing $\xi_i$ in terms of $n_i$ and $\lambda$  involves solving
a cubic equation. Note also that,  first, the left-hand-side of
Eq.~(\ref{var1}) is monotone with respect to $\xi_i$ in the interval
of $0 \le \xi_i < 2$,  and, second,  the expression becomes $1$ as
$\alpha\to 0$. Consider the domain of $2>\xi_i\geq 0$ and $\alpha\ll
(2-\xi_i)$.  Then Eq.~(\ref{var1}) becomes
\begin{eqnarray}
  \xi_i \approx \frac{\alpha \lambda n_i}{2\sqrt{1-\lambda n_i}}, \quad
    h_i \approx 1-\frac{\xi_i}{\lambda n_i} \approx
    1-\frac{\alpha}{2\sqrt{1-\lambda n_i}}.
  \label{var2}
\end{eqnarray}
One observes  that either of the two possibilities is realized at
$\alpha\to 0$: (1)  $\lambda n_i\to 1$ and $\xi_i\to
\mbox{const}\neq 0$; or (2) $\lambda n_i\to\mbox{const}\neq 1$ then
$\xi_i\to 0$. One should also add to these two possible cases  a
third one (that is obviously not explained by Eq.~(\ref{var2}))
where $\xi$ lies exactly on the border of the monotone interval,
i.e. at $\xi=2$ and $h=-1$.

One finds that structurally an instanton consists of three types of
colored bits corresponding to $\xi=0$, $0<\xi<2$ and $\xi=2$
respectively. One also finds that for all the colored bits with
$0<\xi<2$ all the respective $n_i$ approach the same limit, $n_i\to
1/\lambda$, even though respective $\xi_i$ may be distinct. This
results in the following form of the zero a-posteriori
log-likelihood condition at the tree center
\begin{eqnarray}
 \sum_i n_i h_i=N_c-n_*\sum_{i\in {\rm set}}\xi_i-2N_2=0, %\label{zeromag}\\&&
 \quad l_{\rm inst;L}=2m_2+\sum_{i\in {\rm set}}|\xi_i|
 %=2 m_2+\sum_{i\in {\rm set}} \xi_i
 =\frac{N_c-2N_2}{n_*}+2m_2, \label{Qc}
\end{eqnarray}
where ``${\rm set}$'' is the set of colored bits on the original
graph with $0<\xi_i<2$; $N_c$ is the total number of the colored
bits on the computational tree; $N_2$ is the number of marginal,
$\xi=2$, bits on the computational tree; $n_*$ is the number of
replicas found on the computational tree for a bit $i$ that belongs
to the ${\rm set}$ (this is the same number for all $i\in{\rm
set}$); $m_2$ is the number of distinct marginal (i.e. $\xi=2$) bits
on the original graph.

Eq.~(\ref{Qc}) represents the major theoretical result of our
analysis. It explains the rational origin of the effective length
and shows how the effective length depends on the set of integers,
$N_*,N_2,n_*$ and $m_2$ carrying the coding/decoding specific
information. Notice also a remarkable common feature of the
instantons. There is actually a strong degeneracy here if the number
of colored bits with $0<\xi<2$ is two or larger: the effective
length depends only on $\sum_{i\in{\rm set}}\xi_i$, while otherwise,
and modulo the requirement $0<\xi_i<2$, the $\xi_i$ fields can be
chosen arbitrarily. Therefore, to estimate the BER corresponding to
the given set of integers, $N_c,N_2,n_*,m_2$, thus explaining a
continuous family of instantons rather then an individual instanton,
one should also account for the degree of degeneracy (volume of the
respective part of the phase space), $B\sim V(N_c,N_2,n_*,m_2)\times
\exp[-s\cdot l_{\rm inst;L}]$. Our estimate shows that the
$V(N_c,N_2,n_*,m_2)$-terms give sub-leading (i.e. non-exponential in
$s$) corrections to the major (effective length) factors.

\section{Instanton-amoeba calculations in Laplacian channel}
\label{sec:amoeba}

So far we have not discussed how to find the discrete values
$N_c,N_2,n_*$ and $m_2$ that are obviously dependent on the explicit
structure of the LDPC code considered. To solve this problem we
adopt the instanton-amoeba approach of \cite{05SCCV}.

We consider the $(155,64,20)$ LDPC code of \cite{01TSF} as an
example. The minimal Hamming distance of the code is $l_{\rm
inst;L}= 20$, i.e.\ for $s\gg 1$, and if the decoding is ML, the BER
becomes $\sim \exp(-20\cdot s)$ in the case of the Laplacian
channel. The BER is higher, $\sim \exp(-l_{\rm inst;L}\cdot s)$ with
$l_{\rm inst;L}<20$, if one decodes iteratively. For the min-sum
decoding with $4$ iterations we found that three minimal length
instantons are characterized by $l_{\rm a;L} =7.6$, $l_{\rm b;L} =
8$ and $l_{\rm c;L} = 8$ respectively. These instantons were
disclosed as the result of multiple attempts at instanton-amoeba
numerical minimization described above.

The colored parts of the Tanner graph of the code and the respective
parts of the computational tree for the three configurations are
shown in Fig.~1: the three panels (a,b,c) correspond to the three
instantons showing minimal effective lengths. Each panel consists of
two diagrams showing the relevant (colored) part of the Tanner graph
and respective (four iterations deep) part of the computational
tree. Bits/circles are shown with numbers correspondent to the
ordering of the bits explained in \cite{05SCCV}. The shadow bit is
the erroneous one, i.e. it is the bit whose a-posteriori
log-likelihood is exactly zero on the $4$ step of the min-sum
decoding. According to our theoretical analysis detailed above (that
is also confirmed numerically in detail) there are three types of
colored bits/cirlces with $\xi=0$, $0<\xi<2$ and $\xi=2$, shown in
Fig.~1 in white, green and red respectively. The white bits on the
computational tree that are not numbered (and respective bits of the
Tanner graph simply not shown in the Figure) can be chosen
arbitrarily with the only requirement that they are distinct from
any bits shown numbered in the Figure.

\begin{figure}
\includegraphics[width=4in]{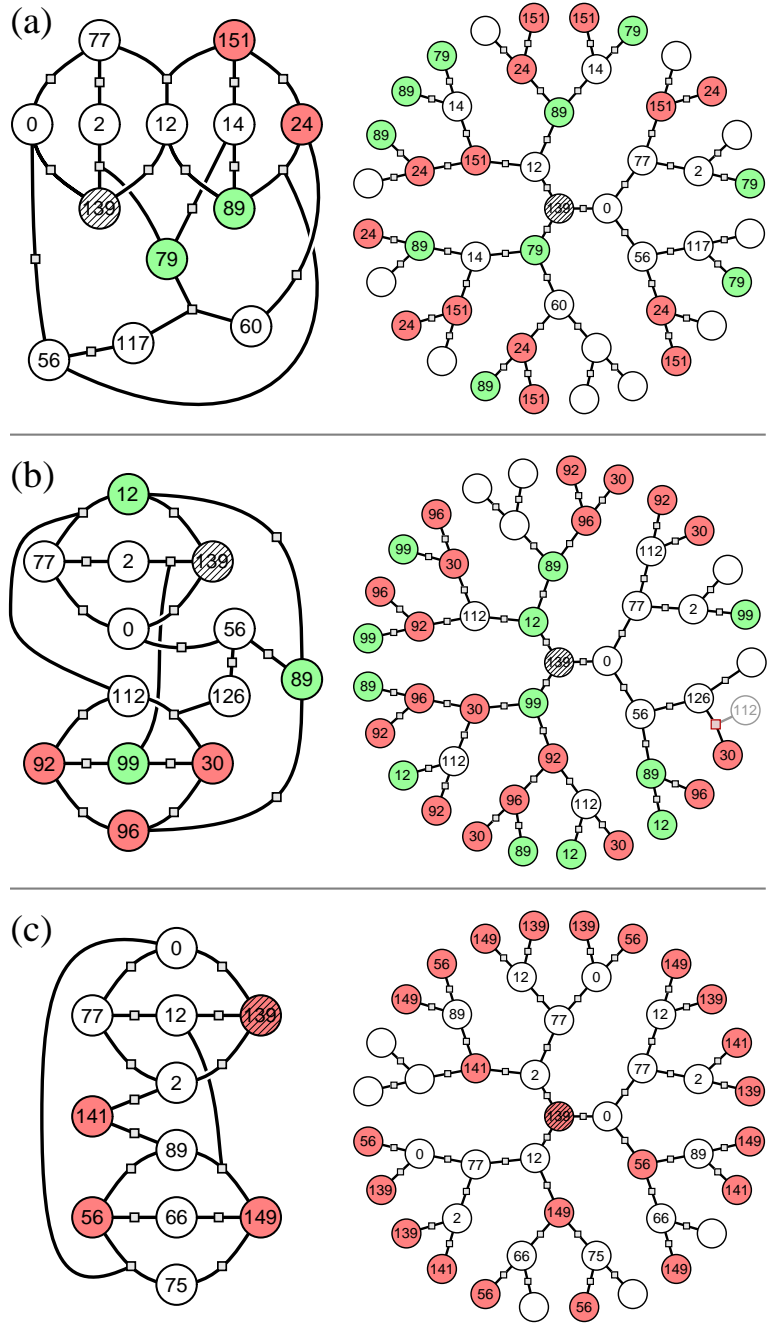}
\caption{ Visualization of three instantons with shortest effective
lengths found by the instanton-amoeba scheme. Only relevant parts of
the Tanner graph for the $(155,64,20)$ code and respective
computational tree with four iterations are shown. Numbers marking
the bits are introduced in accordance with the convention described
in \cite{05SCCV}. The color coding used for bits is white for
$\xi=0$, green for $0<\xi<2$ and red for $\xi=2$. See detailed
explanations in the text.} \label{instanton}
\end{figure}

Configuration (a) shown in Fig.~1a consists of $2$ green bits with
$0<\xi<2$ (numbers $79$ and $89$) each appearing $5$ times on the
computational tree, and $2$ red bits with $\xi=2$ (numbers $24$ and
$151$) each appearing $7$ times on the computational tree. The
remaining $151$ bits carry $\xi=0$ noise. Therefore, the respective
integers corresponding to this instanton are $N_c=46$, $N_2 = 2
\cdot 7 = 14$, $n_* = 5$, $m_2 = 2$, thus resulting according to
Eq.~(\ref{Qc}) in $l_{\rm a;L}=7.6$.  This effective distance was
found with numerical precision by the instanton-amoeba method. The
degeneracy in this family of instantons is one parametric:
$h_3+h_{16}=-8/5$ and $-1\leq h_3,h_{16}\leq 1$. Here, the erroneous
bit, i.e. the bit with zero a-posteriori log-likelihood (marked
striped in the Fig.~1a) is the white one.

Configuration (b) consists of three green bits,  with $0 < \xi < 2$,
appearing four times each on the computational tree, and three red
bits,  with $\xi=2$, appearing $7,6$ and $6$ times respectively. For
this instanton one finds,  $N_2 = 19$, $n_* = 4$, and $m_2 = 3$
($N_c$ is always $46$ for the four iterations decoder) so that
$l_{\rm b;L} = (46 - 2 \cdot 19)/4 + 2 \cdot 3 = 8$. The erroneous
bit (number $139$) is white. One bit, numbered $112$, is special
here. Even though this bit has many replicas on the computational
tree, thus potentially, it would be advantageous to have it green
carrying the non-zero value of the noise,  self-consistency strictly
requires that $\xi_{112}=0$. There are at least two reasons for
this. First of all, if bit $112$ turns green making $\xi_{112} \ne
0$ one of its replicas on the computational tree (the one marked
pale and adjusted to the red check/square in Fig.~1b) screens bit
$30$ (positioned next to the red check) in the sense that this
screened bit will not contribute to $n_{30}$. Consequently, $n_{30}$
becomes $6$ and not $7$, thus making $N_2$ smaller and $l_{\rm
inst;L}$ larger. Second, if $\xi_{112} \ne 0$ then one of the $5$
replicas of the bit $112$ contributes a-posteriori log-likelihood at
the tree center with ``$-$'' sign and $n_{112}$ becomes equal to
$3$, rather than $5$. This number is smaller than $n_* = 4$ thus
leading to the undesired effective length increase. Finally, the
resulting degeneracy in the instanton family is two-parametric
(corresponding to appearance of three, and not four, green bits):
$h_{12}+h_{89}+h_{99}=1$ and $-1\leq h_{12},h_{89},h_{99}\leq 1$.

Configuration (c) has the same effective length as configuration
(b), $l_{\rm c;L}=8$, even though it is very different structurally.
The (c) instanton has no green bits, thus it is non-degenerate and
only one special configuration of $h_i$ fields is realized. There
are four red bits, $56,139,141$ and $149$, appearing on the
computational tree $6,7,4$ and $6$ times respectively. The erroneous
bit,  $139$, is red. The integers are $m_2 = 4$, $N_2 = 23$ and $N_c -
2N_2 = 0$, thus according to Eq.~(\ref{Qc}) $l_{\rm c;L}=2m_2=8$.

\begin{figure}[t]
\includegraphics[width=4in]{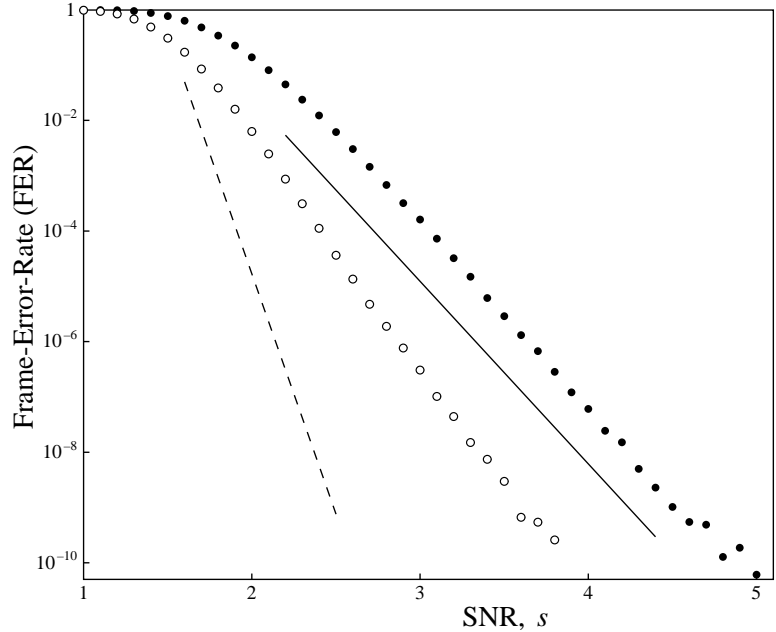}
\caption{Frame-Error-Rate vs Signal-to-Noise ratio, $s$, for the
$(155,64,20)$ code with four iterations of the min-sum decoding over
the Laplacian channel. Solid and dashed lines show slopes (defined
upto a constant shift in the log-lin plot) correspondent to the
instanton with the shortest effective distance, $l_{\rm a;L}=7.6$,
and to the ideal ML decoding with the Hamming distance, $l_{\rm
ML;L}=20$, respectively. Filled and empty circles show results of
direct Monte-Carlo simulations for $4$ and $1024$ iterations
respectively. } \label{ber}
\end{figure}

In \cite{05SCCV} we have argued, following the logic of
\cite{96Wib}, that each instanton is equidistant from some number of
pseudo-codewords, i.e. codewords on the respective computational
tree. This observation is obviously generic and thus it is
applicable as well to the case of the Laplacian channel considered
in the manuscript. For any of the instantons discussed above and
illustrated in Fig.~1 there exists a respective pair of
pseudo-codewords. (As argued in \cite{96Wib} an instanton may be
degenerate --- corresponding to a triple or in principle to even
larger set of pseudo-codewords. This degeneracy, that was found
present in the AWGN channel, was not observed here for the AWLN
channel.) The first pseudo-codeword in a pair is just the all $+1$
codeword ($+1$ sits at every bit). The second pseudo-codeword in a
pair can be introduced according to the following rule: put $-1$ at
any colored (i.e. white, green or red) bit and $+1$ at any uncolored
(thus not shown in Fig.~1) bits. Obviously this choice of the
pseudo-codewords pairs is unambiguous. Indeed the white bits that
are not numbered can be chosen arbitrarily (within appropriate
bounds described above). Moreover, the ambiguity in choosing the
second pseudo-codeword (containing $-1$ bits) is even stronger.
Indeed, this second pseudo-codeword can consists only of two $-1$
bits (all other bits will carry $+1$): one $-1$ should be positioned
at any red bit (with $\xi=2$) from the last generation of the
computational tree and another $-1$ is placed on any uncolored bit
sharing a check with the red bit. This special form of degeneracy is
due to the fact that log-likelihoods assigned to any red bit and to
an uncolored bit adjusted to the red one are the same in absolute
value but opposite in sign.

\section{Validity of instanton approximation at moderate ${\rm SNR}$}
\label{sec:moderate}

The analysis of the previous Section suggests that the leading
$s\to\infty$ asymptotic for BER (and for the Frame-Error-Rate as
well) is governed by the instanton with the lowest effective length
found, i.e. $B\sim \exp[-s\cdot l_{\rm a;L}]$. We have checked this
prediction against direct Monte-Carlo (MC) simulations and found
very good agreement already in the range accessible for the MC,
$B\lesssim 10^{-9}$. See Fig.~2. We observed that the actual
behavior of BER is well described by the instanton not only in the
asymptotic regime of highest SNR but also in the regime of moderate
SNR where there is no a-priori reason to expect the instanton
approximation to work so well. Our further discussion is to
eliminate this point suggesting a plausible explanation for this
surprising generality of the instanton asymptotic.

Let us first clarify why the validity of the instanton asymptotic at
the moderate values of SNR shown in Fig.~2 is surprising. Indeed,
according to Eq.~(\ref{Lap1}) the average value of the noise
configuration length, $l = \sum_i |\xi_i|$, unconstrained by the
requirement to have zero a-posteriori log-likelihood at a bit,  is
$\langle l \rangle \sim N/s$. This means that even at $s = 5$ (the
largest SNR shown in Fig.~2), where the error probability is already
small, ${\rm FER} \sim 10^{-10}$, the typical length of noise
realization is still essentially larger then the respective
instanton prediction: $l \sim 155/5 = 31 > l_{\rm a;L} = 7.6$.
Therefore, naively one expects the instanton to work well at
$s\gtrsim 20$, where $\langle l\rangle\lesssim l_{\rm a;L}$,  while
according to the MC results shown in Fig.~2 the instanton asymptotic
sets already at $s \simeq 2.5$, where ${\rm FER} \sim 10^{-2}$.
(Notice, that the situation for the Gaussian channel is similar, see
Fig.~S2 of \cite{05SCCV}. There the typical $l^2\sum_i \xi_i^2$ is
$\sim N/s^2$, resulting in $<l^2>\approx 30$, while the respective
instanton value is $l_a^2\approx 10.076$.)

Our approach to explaining the validity of the instanton asymptotic
already at the moderate NSR is of the reverse engineering type: we
first formalize what the Monte-Carlo results suggest and then
present a plausible explanation for this phenomenon.

One useful object is the probability distribution function of the
channel noise length (fully unconstrained),  that gets the following
forms for the Gaussian and Laplacian channels respectively: ${\cal
P}_{\rm L}(l) = s^N l^{N-1}[\Gamma(N)]^{-1} \exp(-l \cdot
    s)$  and
${\cal P}_{\rm G}(l) = s^N l^{N-1}2^{1-N/2} [\Gamma(N/2)]^{-1}
    \exp(-l^2 \cdot s^2/2)$.
Typical noise configuration forms a ``spherical'' layer with
``radius'' $l = \sum_i |\xi_i| \approx N/s$ in the $N$-dimensional
noise space (or $l^2 = \sum_i \xi_i^2 \approx N/s^2$ for the
Gaussian channel). The ``spherical" layer becomes thinner as $N$
grows. Obviously ${\cal P}(l)$ does depend on the channel but it
does not depend on the decoding scheme. Another useful object, that
is decoding sensitive, is the (cumulative) distribution function
${\cal F}_{\rm ES}(l)$ of the length for the points positioned
exactly at the error-surface. Estimation for FER is related to the
area of spherical layer that lies outside of the error surface:
${\rm FER} = \int d l \, {\cal P}(l) {\cal F}_{\rm ES}(l)$. Strictly
speaking this relation is exact for the Gaussian channel, but it is
only qualitatively right (up to a $O(1)$ coefficient accounting
accurately for the phase factor) in the Laplacian channel case. The
integral ${\rm FER} = \int d l \, {\cal P}(l) {\cal F}_{\rm ES}(l)$
can be viewed as the Laplace transform. Deducing from the Monte
Carlo simulations that the instanton asymptotic is valid, ${\rm FER}
\propto \exp(-l_{\rm inst;L} \cdot s)$ (or ${\rm FER} \propto
\exp(-l_{\rm inst;G}^2 \cdot s^2/2)$ for the Gaussian channel) one
derives by the inverse Laplace transform ${\cal F}_{\rm ES}(l) = (1
- l_{\rm inst;L}/l)^{N-1}$ (or ${\cal F}_{\rm ES}(l) = (1 - l_{\rm
inst;G}^2/l^2)^{N/2-1}$ for the Gaussian channel).

We suggest that the special dependence of ${\cal F}_{\rm ES}$ on
$l$, deduced from the MC simulations, corresponds to an
$(N-1)$-dimensional area of the part of error-surface with lengths
$l$ or less,  ${\cal F}_{ES}\sim\delta^{N-1}$,  where $\delta$ is
the respective line element. Given that the instanton configuration
is extremal one assumes that, $l-l_{\rm inst}\propto \delta^2$, that
results exactly in the right expression for ${\cal F}_{\rm ES}(l)$
in the Gaussian channel. We ought to assume that in the Laplacian
channel case expansion of $l$ with respect to $\delta\ll l$ about
$l\sim l_{\rm inst}$, is of another type, $l-l_{\rm inst}\propto
\delta$, thus confirming the ${\cal F}_{\rm ES}(l)$ dependence on
$l$ observed in the Laplacian channel simulations.

\section{Conclusions}
\label{sec:concl}

The set of tasks formulated in the introductory Section was accomplished
in Sections \ref{sec:theo}-\ref{sec:amoeba}. We have shown for the
example of the Laplacian channel that the instanton-amoeba optimization
scheme, introduced in \cite{04CCSV} and tested in \cite{05SCCV} for the
Gaussian channel, is computationally efficient. We extended the
theoretical analysis of \cite{05SCCV}, based on the computational tree
approach of \cite{96Wib}, explaining the rational structure of
instantons in the Laplacian channel. The instantons are shown to be
different for different channels considered with the same
coding/decoding scheme. Even though the fact that effective lengths
differ for different channels was already demonstrated in \cite{99FKKR}
for the example of the binary symmetric and binary erasure channels,
this manuscript additionally proved that not only the effective weights
but also the instanton configurations themselves were different
structurally for different channels.

We conclude by noting that the observations made in this manuscript,
extending and complementing our previous works \cite{04CCSV,05SCCV},
virtually solve in a straightforward way the problem of the generic
error-floor analysis. (No sampling in the configurational space, e.g. of
the kind used in \cite{04Ric}, was required). We intentionally choose
the well known $(155,64,20)$ code (used routinely for testing) to
demonstrate the exciting opportunities the instanton-amoeba approach has
to offer. Our next goal is to apply our scheme to a variety of other
(e.g. longer) codes, decoding schemes and channels.

\end{document}